\documentclass[letterpaper,floatfix]{quantumarticle}
\pdfoutput=1
\usepackage[utf8]{inputenc}
\usepackage[english]{babel}
\usepackage[T1]{fontenc}
\usepackage{amsmath,amsfonts,graphicx}
\usepackage{hyperref}
\usepackage[numbers]{natbib}
\usepackage{color}
\usepackage[dvipsnames]{xcolor}
\usepackage{mathtools}
\usepackage{multirow}
\usepackage{tikz}
\usepackage{hyperref}
\hypersetup{colorlinks=true, citecolor=blue, linkcolor=blue}
\usepackage{url}

\makeatletter
\DeclareFontFamily{OMX}{MnSymbolE}{}
\DeclareSymbolFont{MnLargeSymbols}{OMX}{MnSymbolE}{m}{n}
\SetSymbolFont{MnLargeSymbols}{bold}{OMX}{MnSymbolE}{b}{n}
\DeclareFontShape{OMX}{MnSymbolE}{m}{n}{
    <-6>  MnSymbolE5
   <6-7>  MnSymbolE6
   <7-8>  MnSymbolE7
   <8-9>  MnSymbolE8
   <9-10> MnSymbolE9
  <10-12> MnSymbolE10
  <12->   MnSymbolE12
}{}
\DeclareFontShape{OMX}{MnSymbolE}{b}{n}{
    <-6>  MnSymbolE-Bold5
   <6-7>  MnSymbolE-Bold6
   <7-8>  MnSymbolE-Bold7
   <8-9>  MnSymbolE-Bold8
   <9-10> MnSymbolE-Bold9
  <10-12> MnSymbolE-Bold10
  <12->   MnSymbolE-Bold12
}{}

\let\llangle\@undefined
\let\rrangle\@undefined
\DeclareMathDelimiter{\llangle}{\mathopen}%
                     {MnLargeSymbols}{'164}{MnLargeSymbols}{'164}
\DeclareMathDelimiter{\rrangle}{\mathclose}%
                     {MnLargeSymbols}{'171}{MnLargeSymbols}{'171}
\makeatother

\begin{document}

\newcommand{\cE}{\mathcal{E}}
\newcommand{\cL}{\mathcal{L}}
\newcommand{\cB}{\mathcal{B}}
\newcommand{\cH}{\mathcal{H}}
\newcommand{\x}{\mathbf{x}}
\newcommand{\y}{\mathbf{y}}
\newcommand{\s}{\mathbf{s}}

\newcommand{\reals}{\mathbb{R}}

\newcommand{\expect}[1]{\ensuremath{\left\langle#1\right\rangle}}
\newcommand{\ket}[1]{\ensuremath{\left|#1\right\rangle}}
\newcommand{\bra}[1]{\ensuremath{\left\langle#1\right|}}
\newcommand{\braket}[2]{\ensuremath{\left\langle#1|#2\right\rangle}}
\newcommand{\ketbra}[2]{\ket{#1}\!\!\bra{#2}}
\newcommand{\braopket}[3]{\ensuremath{\bra{#1}#2\ket{#3}}}
\newcommand{\proj}[1]{\ketbra{#1}{#1}}
\newcommand{\sket}[1]{\ensuremath{\left|#1\right\rrangle}}
\newcommand{\sbra}[1]{\ensuremath{\left\llangle#1\right|}}
\newcommand{\sbraket}[2]{\ensuremath{\left\llangle#1|#2\right\rrangle}}
\newcommand{\sketbra}[2]{\sket{#1}\!\!\sbra{#2}}
\newcommand{\sbraopket}[3]{\ensuremath{\sbra{#1}#2\sket{#3}}}
\newcommand{\sproj}[1]{\sketbra{#1}{#1}}
\newcommand{\zerovec}{\ensuremath{\vec{\mathbf{0}}}}
\newcommand{\onevec}{\ensuremath{\vec{1}}}

\def\Id{1\!\mathrm{l}}
\newcommand{\Tr}{\mathrm{Tr}}
\newcommand{\kcy}[1]{\textcolor{orange}{[#1]\textsubscript{KCY}}}

\title{Estimating detector error models from syndrome data}

\author{Robin Blume-Kohout}
\affiliation{Quantum Performance Laboratory, Sandia National Laboratories, Albuquerque, NM 87185}
\author{Kevin Young}
\affiliation{Quantum Performance Laboratory, Sandia National Laboratories, Livermore, CA 94550}
\date{April, 2025}

\begin{abstract}
\noindent Protecting quantum information using quantum error correction (QEC) requires repeatedly measuring stabilizers to extract error syndromes that are used to identify and correct errors.  Syndrome extraction data provides information about the processes that cause errors. The collective effects of these processes can be described by a detector error model (DEM).  We show how to estimate probabilities of individual DEM events, and of aggregated classes of DEM events, using data from multiple cycles of syndrome extraction.
\end{abstract}

\maketitle

\section{Introduction}

Fault tolerant quantum computing will be executed on logical qubits protected by quantum error correction (QEC). The heartbeat of a logical quantum register that encodes $k \geq 1$ logical qubits into $n$ physical data qubits is cyclic \textit{syndrome extraction} to measure stabilizers of an error correcting code.  Each syndrome extraction cycle (SEC) is implemented by a quantum circuit on the $n$ data qubits and $n-k$ auxiliary qubits that culminates in parallel measurements of all $n-k$ auxiliary qubits.  Their outcomes comprise an $(n-k)$-bit string of \textit{syndrome bits}, collectively called an \textit{error syndrome} (or just \textit{syndrome}).  The syndromes from $M \geq 1$ consecutive SECs -- optionally including an additional $n$-bit string from a terminating \textit{logical measurement} of all $n$ data qubits -- form what we call a \textit{syndrome history}.  Syndrome histories can be decoded to yield a good guess at how the logical qubit[s] encoded into the data qubits evolved during the $M$ SECs, and therefore how to interpret the terminating logical measurement or how to perform subsequent operations on the encoded qubit[s].

Physical processes cause errors on the $2n-k$ qubits involved in the SECs.  Some of these errors are harmless, some can be detected and identified from the syndrome history, and some are undetectable or misidentified and cause logical errors.  A model for these errors is necessary to decode, correct errors, and protect the encoded information.  A model is more accurate (``better'') if it predicts error events more accurately.  Better models usually reduce the probability of incorrect decoding, which enables lower logical error rates and improves performance of QEC.  For this purpose, a model only needs to predict \textit{observable} phenomena -- i.e., observable patterns and correlations in the syndrome history (which we define to include the terminating $n$-bit logical measurement).  \textit{Detector error models} (DEMs) are excellent models for this purpose (see Section \ref{sec:DEMs} below).  In this paper, we assume that we have syndrome data generated by an unknown DEM, and consider the problem of learning or estimating that DEM from the syndrome data that it generated.

\subsection{Prior work} \label{sec:literature}
Learning and modeling errors in quantum computing systems is a very old and well-studied topic.  Learning error models from syndrome extraction data is a newer idea, initiated (to the best of our knowledge) around 2014 \cite{combes2014situ, fowler2014scalable}.  Notably, Ref.~\cite{fowler2014scalable} proposed the simple strategy of running a decoder on the syndrome data to identify the underlying events (e.g. data qubit flips and/or readout bit flips) in each syndrome history, and estimating each event's probability from its observed frequency.  This approach, subsequently deployed for benchmarking \cite{wootton2022syndrome} and adaptive decoding \cite{huo2017learning}, is limited by \emph{degeneracy}, which occurs when distinct physical events or combinations of events have identical signatures in the syndrome history.  For a simple example, consider an error model in which data qubits 1 and 2 can flip independently, \emph{and} there is a third distinct event that flips both 1 and 2.  We would like to be able to estimate all three events' rates.  But a decoder analyzes each syndrome history independently, one at a time. When it does so, if it sees the signature of ``Data bits 1 and 2 both flip'', it has no way to distinguish coincident flips from correlated flips, and will therefore always decode this signature the same way (either as two coincident events, or a single correlated event).  Accurately estimating the rates of all three events requires some kind of statistical analysis, which may (as in our approach below) not use a decoder at all.

Subsequent work has explored a variety of approaches to this problem.  The most common is to frame it as optimizing decoder performance by adapting the decoder's prior to noise \cite{spitz2018adaptive, nickerson2017analysing, Google2021, chen2021calibrated, sivak2024optimization, darmawan2024optimal, remm2025experimentally,hockings2025improving}.  A complementary approach seeks to estimate a Pauli channel describing the noise in syndrome extraction \cite{wagner2020optimal, wagner2022pauli, wagner2022learning}, sometimes adapting general methods for estimating Pauli channels \cite{flammia2020efficient, rouze2023efficient}, which have also been adapted specifically to the context of syndrome extraction circuits \cite{harper2023learning, hockings2024scalable}.  Ref.~\cite{kobori2024bayesian} proposed and explored a general but strictly numerical approach to estimating custom (non-Pauli) noise models from syndrome data using Bayesian inference.  Several other approaches that defy simple classification \cite{gicev2023quantum, hesner2024using, liepelt2023enhanced} have been explored.

The closest results (and methods) to ours are found in the series of papers by Wagner \emph{et al} \cite{wagner2020optimal, wagner2022pauli, wagner2022learning} that describe the noise in syndrome extraction by a Pauli channel, investigate how well that channel can be learned, and deploy rather sophisticated mathematics to learn as many Pauli error rates as possible.  As far as we can tell, this approach also runs into difficulty because of degeneracy, albeit for a slightly different reason than given above. There are always many Pauli errors that could have caused any observed error syndrome, so a Pauli error model has many unlearnable parameters that complicate the analysis.  We gain a lot of simplification (as well as the ability to deal with readout errors) by framing the problem using DEMs that predict multiple cycles of syndrome extraction.

In that sense, our approach is similar to that of Refs.~\cite{spitz2018adaptive, Google2021, remm2025experimentally}, which describe detector events phenomenologically as we do, and derive very similar formulae.  Our work was motivated in large part by trying to understand exactly what the $p_{ij}$ coefficients \cite{spitz2018adaptive, Google2021} quantify (we answer this below; they are aggregated probabilities of large classes of DEM events).  Ref.~\cite{remm2025experimentally} presents formulae closely related to our results, but without the derivation that we present here.

We expect that our results may be unsurprising to authors of those papers.  With that said, we are not aware of any previous work presenting the full framework that we derive here.

\subsection{Detector error models} \label{sec:DEMs}

A DEM is a statistical model for syndrome histories (including logical measurements).  A syndrome history is a sequence of $M$ $(n-k)$-bit strings (the syndromes for $M$ SECs), plus (optionally) a single $n$-bit string of terminating logical measurement bits.  To maximize simplicity and clarity, we will ignore the terminating measurement data (extending our analysis is left as an exercise for the reader), so our syndrome histories will be $M \times (n-k)$ arrays of bits.

\begin{figure}[t!]
\centering
\includegraphics[width=8cm]{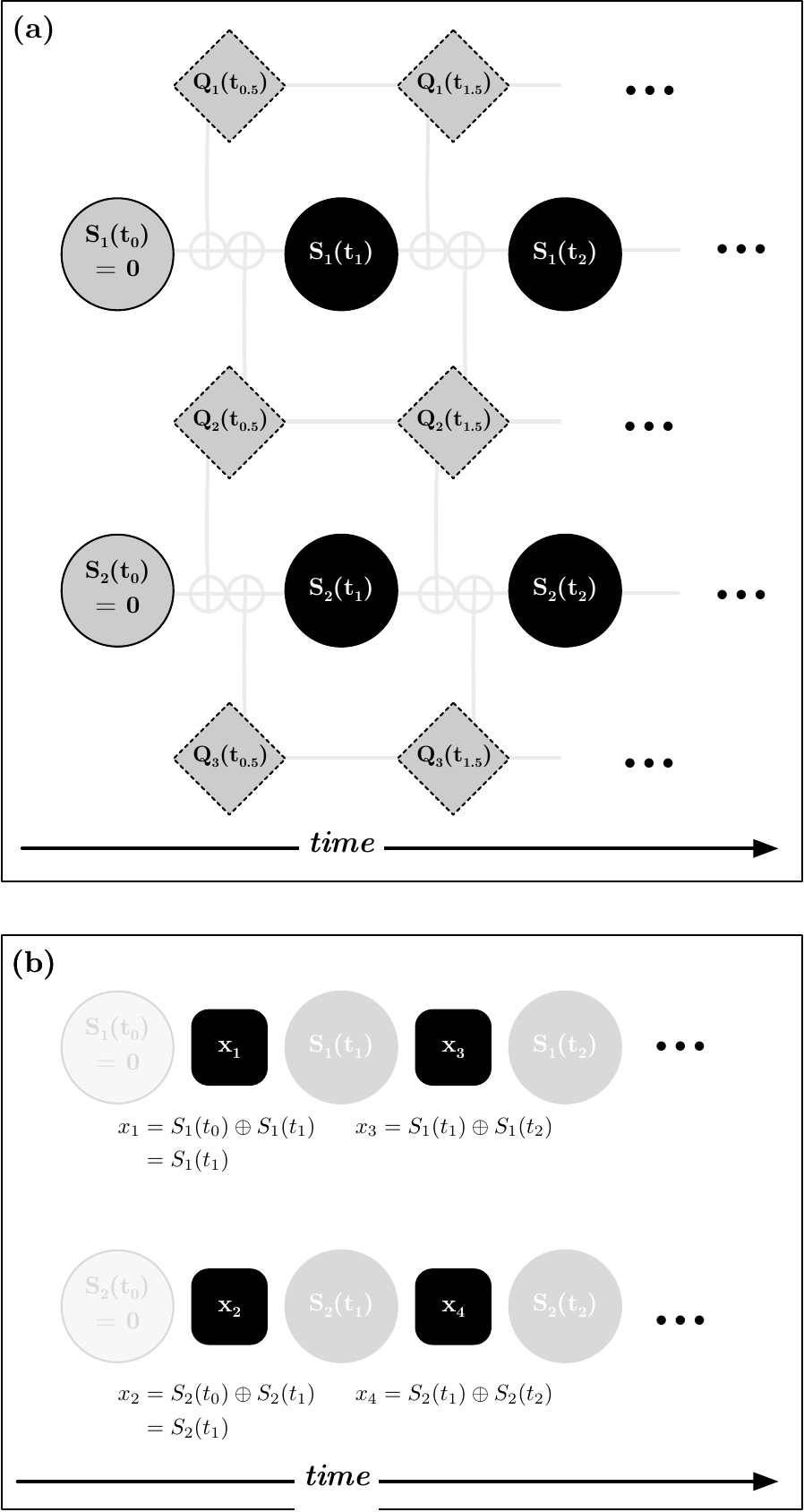}
\caption{Two syndrome extraction cycles (SECs) for a $d=3$ repetition code.  Panel (a) shows the locations where an error could occur on a data qubit $Q_i$ or syndrome readout $S_j$, and panel (b) shows the 4 detector bits, each of which corresponds to a parity constraint on (i.e., an XOR between) consecutive syndrome bits.  The ``$t=0$'' values of the syndrome bits $S_1$ and $S_2$ are not measured, nor are they locations where errors can occur; they are defined as $0$, so that detectors $x_1$ and $x_2$ detect errors in the initialization of the syndrome bits.}
\label{fig:d3detectors}
\end{figure}

Errors on data qubits cause \textit{persistent} changes in the values of syndrome bits.  The first step in decoding is therefore to identify change events -- i.e., points in time where one or more syndrome bits change.  The parity (XOR) of two consecutive syndrome measurements detects such change events, and is called a \textit{detector}\footnote{In general, detectors are simply constraints on the parity of certain sets of measurement bits that are guaranteed to hold when the circuit is run without error.}.  Following common practice, we transform raw syndrome histories into \textit{detector histories} by computing the XOR between every consecutive pair of syndrome bits.  Syndrome bits usually\footnote{If the system was initialized into a code state, all syndrome bits are zero at $t=0$.  However, syndrome extraction is sometimes used to initialize the system into the code.  If ``initializing'' SECs are included in the syndrome history, then their results replace \textit{a priori} values.} have a defined value prior to any measurements (at $t=0$, see Fig.~\ref{fig:d3detectors}). Combining these \textit{a priori} values with $M$ cycles of measured syndromes yields syndrome values at $M+1$ times.  The XOR transforms each $(M+1) \times (n-k)$ syndrome history into an $M\times (n-k)$ detector history.  The shape of these histories is important for designing decoders, but not for our purposes.  Therefore, we treat detector histories as $N$-bit strings with $N = M\times (n-k)$.  Hereafter, we do not use $n$, $k$, or $M$ as defined in this section, and in fact we use these symbols for other purposes. $N$ will continue to represent the length of a syndrome bit string. 

If no errors at all occur, then every syndrome bit will be read out accurately, and its value will never change.  Therefore, every detector bit will be 0 in every history.  If an error \textit{does} happen during a SEC, it will flip some number of detector bits from 0 to 1.  We call the subset of detector bits flipped by an error its \textit{support}.  An error's support may be distributed over multiple time indices.  For example, readout errors cause a syndrome bit to be read incorrectly for just one SEC, and this flips \textit{two} detector bits, at consecutive times.  If two errors occur, and their supports overlap, then the detector bits in the overlap will be flipped \textit{twice} -- once by each error -- and end up reading 0 again.  These observations yield rules that define DEMs:
\begin{enumerate}
\item Errors are independent events that can occur during syndrome extraction.
\item Each error flips a subset of detector bits (a \textit{DEM event}) and is uniquely defined by this subset.
\item If two errors flip exactly the same subset of detector bits, then they are indistinguishable and we treat them as the same event.
\end{enumerate}
A DEM is therefore a list of DEM events ($E$) and their probabilities ($p_{E}$):
\begin{equation}
\label{eq:DEMdef}
    \mathrm{DEM} = \{(E_i, p_{E_i}): i = 1\ldots L\}.
\end{equation}
A DEM event can be represented equally well in two ways:
\begin{enumerate}
    \item as a subset of detector bits, e.g.~$E = \{1,4,5\}$ is an event that flips the first, fourth, and fifth bits of the detector history.
    \item as an $N$-bit binary string, e.g.~$\s = \mathtt{[10011000]}$ for $N=8$ represents the same event.
\end{enumerate}
We will use these notations interchangeably, but we use $\s$ more frequently.

Here is a sample DEM for $M=2$ rounds of a $k=3$-bit repetition code (see Fig.~\ref{fig:d3detectors}).  There are $N = M\times(k-1)=4$ detector bits.  The first two bits in $\s = [x_1x_2x_3x_4]$ represent the detectors from the first SEC, followed by those for the second SEC.  For clarity, each event is labeled by its cause, e.g. ``$Q_1(t_{1.5})$'' indicates a bitflip error on data qubit 1 between $t=1$ and $t=2$, while ``$S_1(t_1)$'' indicates incorrect readout of syndrome bit 1 at $t=1$:

\begin{center}
\begin{tabular}{|c|c|c|c|}
 \hline
 \multicolumn{4}{|c|}{Sample DEM} \\
 \hline
 Event $E$ & $\s$ & Probability & Cause\\
 \hline
\{1\} & $\texttt{[1000]}$ & $p_{[1000]}$ &  $Q_1(t_{0.5})$ \\
\{1,2\} & $\texttt{[1100]}$ & $p_{[1100]}$ &  $Q_2(t_{0.5})$ \\
\{2\} & $\texttt{[0100]}$ & $p_{[0100]}$ &  $Q_3(t_{0.5})$ \\
\{1,3\} & $\texttt{[1010]}$ & $p_{[1010]}$ &  $S_1(t_1)$ \\
\{2,4\} & $\texttt{[0101]}$ & $p_{[0101]}$ &  $S_2(t_1)$ \\
\{3\} & $\texttt{[0010]}$ & $p_{[0010]}$ &  $Q_1(t_{1.5}) \mathrm{\ or\ } S_1(t_2)$ \\
\{3,4\} & $\texttt{[0011]}$ & $p_{[0011]}$ &  $Q_2(t_{1.5})$ \\
\{4\} & $\texttt{[0001]}$ & $p_{[0001]}$ &  $Q_3(t_{1.5}) \mathrm{\ or\ } S_2(t_2)$\\ 
 \hline
\end{tabular}
\end{center}

A DEM defines a probability distribution over detector histories ($N$-bit strings), which can be represented as a $2^N$-element vector $\vec P = \sum_\x \mathrm{Pr}(\x) \vec{\x}$, where $\vec{\x}$ indicates the unit vector in $\mathbb{R}^{2^N}$ indexed by the $N$-bit detector string $\x$. It is easy to generate a sample from this distribution as follows:
\begin{enumerate}
    \item Initialize an error-free syndrome history, i.e.~an $N$-bit string of 0s, $\x = \mathtt{[0^N]}$.
    \item For each DEM event $(\s, p_\s)$, add $\s$ to $\x$ (mod 2) with probability $p_\s$.  That is, generate a random bit from the distribution $[1-p_\s,p_\s]$, and iff that bit is 1, map $\x \to \x \oplus \s$.
\end{enumerate}
An equivalent construction is:
\begin{enumerate}
    \item Construct a binary $N\times L$ matrix $\mathcal{E}$ whose $i$th column is $\s_i$.
    \item Construct an $L$-bit binary vector $\mathbf{q}$ whose $i$th bit is 1 with probability $p_{\s_i}$ and 0 with probability $1-p_{\s_i}$.
    \item Multiply them to obtain $\x = \mathcal{E}\mathbf{q}$.
\end{enumerate}
If $2^N$ is not too big, then $\vec{P}$ can be constructed explicitly as follows:
\begin{enumerate}
    \item Initialize the probability distribution associated with an error-free syndrome history, i.e., the $2^N$-dimensional vector $ \vec P_0 = \vec{\mathbf{0}} = (1,0,0, \cdots)$. 
    \item For each DEM event $(\s, p_\s)$, multiply the probability vector by the stochastic matrix \begin{equation} L_\s = (1-p_\s)\Id + p_\s X_\s. \label{eq:stochasticmatrix} \end{equation}
    where $X_\s = \bigotimes_i X^{s_i}$ is the operator that flips all detector bits in the support of $\s$ and $X$ is the usual Pauli $X$ matrix.
\end{enumerate}
Iterating over all DEM events produces the distribution
\begin{equation}\label{eq:pfinal}
    \vec P = \left[\prod_\s L_\s \right] \vec P_\mathrm{0}.
\end{equation}
This representation is impractical for calculation when $N$ is even moderately large---it requires storing and processing a $2^N$-dimensional vector---but will be useful for deriving our results. 

In this work we consider the \textit{inverse} problem: given some data (detector histories) sampled from an unknown DEM, what events have non-zero probability, and what are their probabilities?  Previous work \cite{wagner2020optimal, wagner2022pauli, wagner2022learning} considered the closely related question of estimating Pauli channels from syndrome data, but degeneracy means that Pauli channels can't generally be estimated uniquely from data.  Detector error models are essentially Pauli channels mod stabilizers -- which is exactly what \textit{can} be estimated uniquely from data.  We solve this problem using generalized Fourier analysis \cite{merkel2021randomized}, specifically the Walsh-Hadamard transform.  This approach will be familiar and unsurprising to readers who have worked on Pauli noise estimation \cite{flammia2020efficient, rouze2023efficient,harper2023learning, hockings2024scalable}.

\section{Inferring DEMs from data}
\label{sec:panda}

In Ref.~\cite{Google2021}, Google Quantum AI showed how to learn some interesting things about the DEM from data.  They used empirical data to estimate moments of the probability distribution over detector bits $\x$ -- e.g. $\expect{x_i}$ or $\expect{x_ix_j}$ -- and from those moments they estimated properties of the DEM probabilities $\{p_{E}\}$.  This relationship was originally derived and used by Spitz \textit{et al} in Ref.~\cite{spitz2018adaptive}.  It was very recently generalized by Remm \textit{et al} in Ref.~\cite{remm2025experimentally}, but without a derivation.  We provide that derivation, via a useful change of variables from detector bit moments to \textit{polarizations} and from DEM event probabilities to \textit{attenuations}.

We motivate this change of variables by observing that the stochastic matrices $L_\mathrm{s}$ in Eqs.~\ref{eq:stochasticmatrix}-\ref{eq:pfinal} all commute with each other, and can be simultaneously diagonalized by a Walsh-Hadamard transform, $H$, defined\footnote{The matrix form of the $N$-bit Walsh-Hadamard transform, $H$, is the $N$th tensor power of the $2\times 2$ Hadamard matrix, $H=H_2^{\otimes N}$.} by
\begin{equation}
    H \vec{\s} = \frac{1}{2^{N/2}}\sum_\y (-1)^{\y\cdot\s} \vec{\y}.
\end{equation}
$H$ is self-inverse.  Applying this transformation to the formula for $\vec{P}$ in Eq.~\ref{eq:pfinal} gives:
\begin{align}
    H \vec P &= \left[\prod_\s H L_{\s} H\right] H \vec P_\mathrm{0},
\end{align}
with
\begin{equation}
    H L_\s H = (1-p_\s)\Id + p_\s Z_\s.
    \label{eq:transformedL}
\end{equation}
Here $Z_\s = \bigotimes_i Z^{s_i}$ and $Z$ is the usual Pauli $Z$ matrix. In this representation, the matrices ($H L_\s H$) are diagonal, which simplifies analysis fantastically.  The price for this simplification is describing the data not by probabilities (elements of $\vec{P}$) but by \textit{polarizations} (elements of $H\vec{P}$).

\subsection{Parities and Polarizations} \label{sec:polarization}

The elements of $H\vec P$ can be computed directly as:
\begin{align}
\label{eq:hp}
\begin{split}
        H \vec P 
        &= H \sum_x \Pr(\x) \vec \x \\
        &= \frac{1}{2^{N/2}}\sum_{\x,\y} \Pr(\x) (-1)^{\x\cdot\y}  \vec\y \\
        &= \frac{1}{2^{N/2}} \sum_{\y} \expect{(-1)^{\x\cdot\y}}_\x \vec\y \\
        & \equiv \frac{1}{2^{N/2}} \sum_\y \expect{z_\y} \vec\y \equiv \frac{1}{2^{N/2}}\expect{\vec{z}},
\end{split}
\end{align}
where we have defined $z_{\y} = (-1)^{\x\cdot\y}$ and $\vec{z} = \sum_{\y}{z_{\y}\vec{\y}}$.  We call $z_\y$ a \textit{parity}, and $\expect{z_\y}$ a \textit{polarization}.  Expectations are almost always taken over detector histories, and so we drop the subscript, writing $\expect{\cdot}$ rather than $\expect{\cdot}_\x$ except in the rare case of expectations over a different variable.

A polarization is the expected value of a parity.  The parity of a single detector bit $x_i$ is simply $z_i = (-1)^{x_i} = 1-2x_i$, which equals $1$ if $x_i=0$ and $-1$ if $x_i=1$.  If $x_i$ is a random variable, then its parity $z_i$ is also a random variable, but its polarization has a well-defined value $\expect{z_i} = \expect{(-1)^{x_i}} = 1-2\expect{x_i}$.

We will mostly be concerned with parities and polarizations of \textit{multiple} bits, e.g. $z_iz_j = (-1)^{x_i \oplus x_j}$.  Multi-bit polarizations can be nontrivial functions of the corresponding bit expectation values.  For example,
\begin{align}
    \expect{z_iz_j} 
    &= \expect{(1-2x_i)(1-2x_j)} \\
    &= 1-2\expect{x_i}-2\expect{x_i}+4\expect{x_ix_j},
\end{align}
which careful readers of Ref.~\cite{Google2021} may recognize.  Whereas multi-bit random variables like $x_ix_j$ are unbalanced functions of the component bits -- e.g., $x_ix_j$ equals 1 if $\x = [11]$, or 0 if $\x\in\{\mathtt{[00],[01],[10]}\}$ --  multi-bit polarizations are balanced.  

Instead of writing multi-bit parities as products of single-bit parities, we index them by $N$-bit strings $\y$. In Eq.~\ref{eq:hp} we defined
\begin{equation}
    z_{\y} = (-1)^{\x\cdot\y},
\end{equation}
where $\y$ is an $N$-bit string, and $\x\cdot\y = (\sum_i{x_iy_i}\ \mathrm{mod}\ 2)$.  In this notation, single bit parities like $z_3$ are written as (e.g., for $N=8$)
\begin{equation}
    z_3 \equiv z_{\mathtt{[00100000]}}.
\end{equation}
In general, $z_{\y_1}z_{\y_2} = z_{\y_1\oplus \y_2}$.  

Given a sample of $K$ $N$-bit strings $\{\x_i\}$, the empirical value of any polarization can be computed efficiently as
\begin{equation} \label{eq:EmpiricalPolarization}
    \widehat{\expect{z_{\y}}} = \frac{1}{K}\sum_{i=1}^K{(-1)^{\mathbf{x_i}\cdot\y}}.
\end{equation}
This is the \textit{sample polarization}, a good estimator of the true polarization $\expect{z_\y}$ that is determined by $\vec{P}$.

We will make \textit{extensive} use of the negative logarithm of a polarization. We call this a \underline{de}polarization and denote it by
\begin{equation}
    \omega_\y \equiv -\ln(\expect{z_{\y}}). \label{eq:depolarization}
\end{equation}

\subsubsection{Error bars and uncertainties}

Since Eq.~\ref{eq:EmpiricalPolarization} can be used to estimate polarizations from data, a very brief word on uncertainty quantification is warranted.  In the remainder of this paper, we'll generally focus on how to calculate DEM properties from \textit{known} polarizations.  But for all practical purposes, it's important to keep in mind that these $\expect{z_\y}$ and $\omega_\y$ are only \textit{estimated} from $K$ samples.  Estimates fluctuate around the truth due to finite-sample fluctuations, so estimated quantities should be qualified by uncertainties, a.k.a.~\textit{error bars}.  Serious statisticians may prefer confidence or credible intervals/regions.  We defer serious statistics to future work.

The uncertainty or error bar for a single estimated polarization $\expect{z_\y}$ can reasonably be estimated by the standard error of its mean, $\sigma / \sqrt{K}$, where $\sigma^2$ is the [sample] variance of $z_\y$.  The variance of $z_\y$ is surprisingly simple:
\begin{equation}
    \sigma^2(z_\y) = \expect{z_\y z_\y} - \expect{z_\y}^2 = 1-\expect{z_\y}^2,    
\end{equation}
so reasonable error bars for a single $\expect{z_\y}$ are
\begin{equation}
    \sigma_{\expect{z_\y}} = \frac{1-\expect{z_\y}^2}{\sqrt{K}}.
\end{equation}
We make more use of depolarizations (Eq.~\ref{eq:depolarization}), and the easiest way to propagate uncertainty in $\expect{z_\y}$ to uncertainty in $\omega_\y$ is by multiplying by the absolute value of the Jacobian to get
\begin{equation}
    \sigma_{\omega_\y} = \frac{1-\expect{z_\y}^2}{\expect{z_\y}\sqrt{K}},
\end{equation}
although it should be noted that whenever $\sigma_{\omega_\y}\geq1$, that means $\sigma_{\expect{z_\y}} \geq \expect{z_\y}$ and so $\sigma_{\omega_\y}$ is effectively infinite because $\expect{z_\y}$ could reasonably be zero.

Estimates of DEM properties almost always depend on \textit{multiple} polarizations.  They are not independent.  To correctly propagate uncertainty, it is necessary to treat them as a vector $\expect{\vec{z}}$, and compute their covariance matrix.  Its elements are simple to calculate:
\begin{align}
    \mathrm{cov}(z_{\mathbf{y}},z_{\mathbf{y'}}) &= \expect{z_{\mathbf{y}} z_{\mathbf{y'}}} - \expect{z_{\mathbf{y}}}\expect{z_{\mathbf{y'}}} \\
    &= \expect{z_{\mathbf{y}\oplus\mathbf{y'}}} - \expect{z_{\mathbf{y}}}\expect{z_{\mathbf{y'}}}.
\end{align}
This covariance matrix can be propagated through calculations to yield estimated error bars on (and, if necessary, covariances between) estimated DEM properties.  However, in practice it may be easier to use resampling techniques such as the bootstrap or the jackknife \cite{shao2012jackknife}.  We defer detailed analysis of these techniques to future work, but since statistical significance testing will probably be important to reconstructing DEMs accurately, a brief discussion of the options seemed warranted.

\subsection{Attenuation}

The transition matrices $L_\s$ in Eq.~\ref{eq:transformedL} are diagonal in the basis of polarizations (i.e., the Walsh-Hadamard transform diagonalizes them).  Each one has two degenerate eigenvalues.  The eigenspace $\mathsf{span}\left(\left\{\y \,\vert\, \s\cdot\y=0 \right\}\right)$ is associated with the eigenvalue $1$, and $\mathsf{span}\left(\left\{\y \,\vert\, \s\cdot\y=1 \right\}\right)$ with the eigenvalue $1-2p_\mathrm{s}$. 

This can be interpreted as follows. A DEM event $\s$ flips every bit in the support of $\s$ if it occurs.  That flips the sign of every parity $z_{\y}$ for which $\y$ shares an odd number of 1 bits with $\s$.  This occurs with probability $p_{\s}$. So the effect of $L_{\s}$ is to attenuate each polarization $\expect{z_\y}$ for which $\y$ shares an odd number of 1 bits with $\s$, by a factor $(1-2p_\s)$:
\begin{equation}
    \expect{z_\y} \to (1-2p_\s)\expect{z_\y}.
\end{equation}
We call
\begin{equation}
d_\s = (1-2p_\s)
\end{equation}
the \textit{decay factor} of event $\s$, and we call the decay factor's negative logarithm
\begin{equation}
    a_\s = -\ln(d_\s) \approx 2p_\s + O(p_\s^2)
\end{equation}
the \textit{attenuation} of event $\s$.  

\subsubsection{Reduced DEMs}

A detector error model is a list of independent DEM events $(E)$ with probabilities $(p_E)$ (Eq.~\ref{eq:DEMdef}).  A DEM event $E$ is defined by the set of detectors that it flips.  As we saw above, it's often useful to represent each event as an $N$-bit string $\s$.  But sometimes it's useful to ignore some detectors, restricting our attention to a $k$-bit subset of the $N$ detector bits.  The original DEM's action on this subset of $k$ detectors is described by a \emph{reduced DEM}.

We construct the reduced DEM by mapping each $N$-bit event label in the original DEM to its $k$-bit substring on the bits of interest.  However, when we do this, there may be collisions---events that had distinct actions on $N$ detector bits, and distinct probabilities, but flip exactly the same subset of the $k$ bits of interest.  Since these events are now indistinguishable (and will be labeled by the same $k$-bit string in the reduced DEM), we must aggregate their probabilities.

Interestingly, the rule for aggregating probabilities is not simple addition, but rather \textit{exclusive} addition. If two $N$-bit events $A$ and $B$ both map to the same $k$-bit event $C$, then $C$ occurs \textit{only} if just one of $A$ or $B$ happens. Their aggregated probability is $p_{C} = p_A(1-p_B) + (1-p_A)p_B$. This rule can be applied iteratively to combine an arbitrary number of identical events. It leads to ugly expressions, though.

Happily, the math gets much simpler if we use attenuation instead of probability. Recall that the attenuation is defined $a = -\ln(1-2p)$.  Straightforward arithmetic shows that
\begin{equation}
    \begin{split}
        a_{C} &= -\ln\left(1-2p_{C}\right) \\
            &= -\ln\left(1-2(p_A(1-p_B) + (1-p_A)p_B)\right) \\
            &= -\ln\left(1-2p_A -2p_B +4 p_A p_B \right) \\
            &= -\ln\left((1-2p_A)(1-2p_B)\right) \\
            &= a_A + a_B.
    \end{split}
\end{equation}
This nice property -- that attenuations, unlike probabilities, \textit{do} aggregate linearly -- motivates the use of $a$ rather than $p$ to quantify the strength of a DEM event.

Here is a worked example.  Consider the following 4-bit DEM:
\begin{center}
\begin{tabular}{|c|c|c|}
 \hline
 \multicolumn{3}{|c|}{$N=4$ DEM} \\
 \hline
 Event ($\s$) & Probability & Attenuation \\
 \hline
 \texttt{[1100]} & $p_{\{1,2\}}$  & $a_{\{1,2\}}$ \\
 \texttt{[1010]} & $p_{\{1,3\}}$  & $a_{\{1,3\}}$ \\
 \texttt{[0110]} & $p_{\{2,3\}}$  & $a_{\{2,3\}}$ \\
 \texttt{[1001]} & $p_{\{1,4\}}$  & $a_{\{1,4\}}$\\
 \texttt{[0101]} & $p_{\{2,4\}}$  & $a_{\{2,4\}}$ \\
 \texttt{[0011]} & $p_{\{3,4\}}$  & $a_{\{3,4\}}$ \\
 \hline
\end{tabular}
\end{center}
We can construct a reduced DEM for the first two bits.  The event \texttt{[0011]} disappears entirely, since it has no effect on $x_1$ or $x_2$, and the other five 4-bit events get aggregated into just three 2-bit events.  We can compute these three events' probabilities,
\begin{center}
\begin{tabular}{|c|c|}
 \hline
 \multicolumn{2}{|c|}{ $k=2$ reduced DEM for $\{x_1,x_2\}$} \\
 \hline
 Event ($\s$) & Probability \\
 \hline
 \texttt{[11]} & $p^{\rm r}_{\{1,2\}} \equiv p_{\{1,2\}}$  \\
 \texttt{[10]} & $p^{\rm r}_{\{1\}} \equiv p_{\{1,3\}}+p_{\{1,4\}} - 2p_{\{1,3\}}p_{\{1,4\}}$ \\
 \texttt{[01]} & $p^{\rm r}_{\{2\}} \equiv p_{\{2,3\}}+p_{\{2,4\}} - 2p_{\{2,3\}}p_{\{2,4\}}$ \\
 \hline
\end{tabular}
\end{center}
but it's much simpler to use attenuations instead:
\begin{center}
\begin{tabular}{|c|c|}
 \hline
 \multicolumn{2}{|c|}{ $k=2$ reduced DEM for $\{x_1,x_2\}$} \\
 \hline
 Event ($\s$) & Attenuation \\
 \hline
 \texttt{[11]} & $a^{\rm r}_{\{1,2\}} \equiv a_{\{1,2\}}$ \\
 \texttt{[10]} & $a^{\rm r}_{\{1\}} \equiv a_{\{1,3\}} + a_{\{1,4\}}$ \\
 \texttt{[01]}  & $a^{\rm r}_{\{2\}} \equiv a_{\{2,3\}} + a_{\{2,4\}}$ \\
 \hline
\end{tabular}
\end{center}
This example illustrates two useful points.  First, attenuations are clearly easier to deal with than probabilities.  Second, notation can be a bit tricky.  We chose to index event probabilities by sets of detectors (e.g.~$\{1,4\}$) instead of strings (e.g.~\texttt{[1001]}) because we find the math to be more transparent -- e.g., obviously when bits 3 and 4 are ignored, $\{1,3\}$ and $\{1,4\}$ both reduce to $\{1\}$.  But doing so required distinguishing reduced-DEM probabilities (e.g.~$p^{\rm r}_{\{1,2\}}$) from their original-DEM counterparts (e.g.~$p_{\{1,2\}}$) to avoid possible collisions, because the original DEM could have also had a $p_{\{1\}}$ that wouldn't equal $p^{\rm r}_{\{1\}}$.

\subsection{Polarizations are determined by attenuations}

DEM events attenuate polarizations and increase depolarizations.  The value of any given polarization $\expect{z_\y}$ is determined by the attenuations of all the DEM events $\s$ for which $\y\cdot\s = 1$:
\begin{align}
    \expect{z_\y} &= \prod_{\s}{ d_\s^{\y\cdot \s}}
\end{align}
Taking the negative log of this equation gives the following linear relationship between depolarizations and attenuations:
\begin{align}
    \omega_\y &= \sum_{\s} (\y\cdot \s)\, a_\s . \label{eq:omega}
\end{align}
This relationship can be inverted to learn or infer attenuations --- which define the DEM --- from empirical depolarizations computed from data using Eq.~\ref{eq:EmpiricalPolarization}.

\subsection{Inferring attenuations from polarizations}

Eq.~\ref{eq:omega} is a linear equation for $\omega_\y$.  There are $2^N$ possible depolarizations $\omega_\y$ (labeled by parities $\y$), and there are $2^N$ possible attenuations $a_\s$ (labeled by DEM events $\s$), with $a_\s=0$ if $\s$ does not appear in the DEM.  If we arrange all the $\omega_\y$ into a $2^N$-dimensional vector $\vec{\omega}$, and arrange all the $a_\s$ into a $2^N$-dimensional vector $\vec{a}$, then we can write Eq.~\ref{eq:omega} as
\begin{equation}
    \vec{\omega} = W\vec{a},\ \mathrm{where}\ W_{\y,\s} = \y\cdot\s. \label{eq:matrix}
\end{equation}
The matrix $W$ is a binary matrix whose rows indicate which attenuations affect each polarization. It is closely related to the Walsh-Hadamard transformation matrix $H$, which is self-inverse ($H^2=\Id)$.  More precisely, 
\begin{equation}
    W = \frac12\left(\vec{1}\vec{1}^T - 2^{N/2}H\right),
\end{equation}
where $\vec{1}$ is the $2^N$-dimensional vector of 1s.  

We could solve Eq.~\ref{eq:matrix} for $\vec{a}$ by inverting $W$.  However, $W$ is not actually invertible, because it annihilates the vector $\zerovec = 2^{-N/2}H\vec{1} = (1,0,0,\ldots)$, which is supported entirely on $\y_0 = \mathtt{[0^N]}$:
\begin{align}
\begin{split}
    W\vec{\mathbf{0}} &= \frac12\left(\vec{1}\vec{1}^T - 2^{N/2}H\right)\vec{\mathbf{0}} \\
    &= \frac12\left(\vec{1} - \vec{1}\right) = 0.
\end{split}
\end{align}
Taking the transpose we also have $\vec{\mathbf{0}}^T W = 0$. Fortunately, this vector is irrelevant, because both $\omega_{\y_0}$ and $a_{\y_0}$ are always zero.  It is sufficient to invert $W$ on its $(2^N-1)$-dimensional support.  This is accomplished by the Moore-Penrose pseudoinverse:
\begin{align}
    W^+ &= -\frac{2}{2^{N/2}} \Pi_{\zerovec^\perp} H \,\Pi_{\zerovec^\perp}  \label{eq:mpinv}\\
        &= \frac{2}{2^N} \left(\vec{\mathbf{0}}\vec{1}^T + \vec{1}\vec{\mathbf{0}}^T - \vec{\mathbf{0}}\vec{\mathbf{0}}^T - 2^{N/2}H\right),
\end{align}
where $\Pi_{\zerovec^\perp} =  \Id - \zerovec\zerovec^T$ is the projector onto the orthogonal complement of $\zerovec$. The $W^+$ matrix is proportional to the usual Walsh-Hadamard matrix, but with the first row and first column set to zero. We can demonstrate that $W^+$ acts as an inverse on the support of $W$ directly: 
\begin{align}
\begin{split}
W^+ W &= \frac{1}{2^N}\left(\vec{\mathbf{0}}\vec{1}^T - 2^{N/2}H\right)\left(\vec{1}\vec{1}^T - 2^{N/2}H\right)\\
    &= \frac{1}{2^N}\left(2^N \vec{\mathbf{0}}\vec{1}^T - 2^{N}\vec{\mathbf{0}}\vec{1}^T -2^N \vec{\mathbf{0}} \vec{\mathbf{0}}^T + 2^{N} \Id \right)\\
    &= \Id - \vec{\mathbf{0}}\vec{\mathbf{0}}^T.  
\end{split}
\end{align}
$W^+$ can then be used to solve Eq.~\ref{eq:matrix} as $\vec{a} = W^+\vec{\omega}$. And as long as we restrict $\vec a$ and $\vec \omega$ to the support of $W$, then we can drop the projectors in Eq.~\ref{eq:mpinv} and simply use the scaled Walsh-Hadamard transformation directly as an inverse. Solving Eq.~\ref{eq:matrix} therefore gives:
\begin{equation}
    \vec{a} = -\frac{2}{2^{N/2}}H\vec{\omega}. \label{eq:result1}
\end{equation}
This simple equation is our first main result.  Unpacked, it gives explicit formulae for the attenuation, decay factor, or probability of any DEM event $\s$ in terms of observable properties (polarizations or depolarizations) of syndrome data:
\begin{align}
    a_\s &= -\frac{2}{2^{N}}\sum_\y{(-1)^{\y\cdot\s}\omega_{\y}}, \label{eq:result2}\\
    d_\s &= \left(\prod_\y{\expect{z_\y}^{(-1)^{\y\cdot\s\oplus 1}}}\right)^{2^{1-N}}, \label{eq:result3} \\
    p_\s &= \frac12 - \frac12\left(\prod_\y{\expect{z_\y}^{(-1)^{\y\cdot\s\oplus 1}}}\right)^{2^{1-N}}. \label{eq:result4}
\end{align}

\begin{figure}[t!]
\centering
\includegraphics[width=7cm]{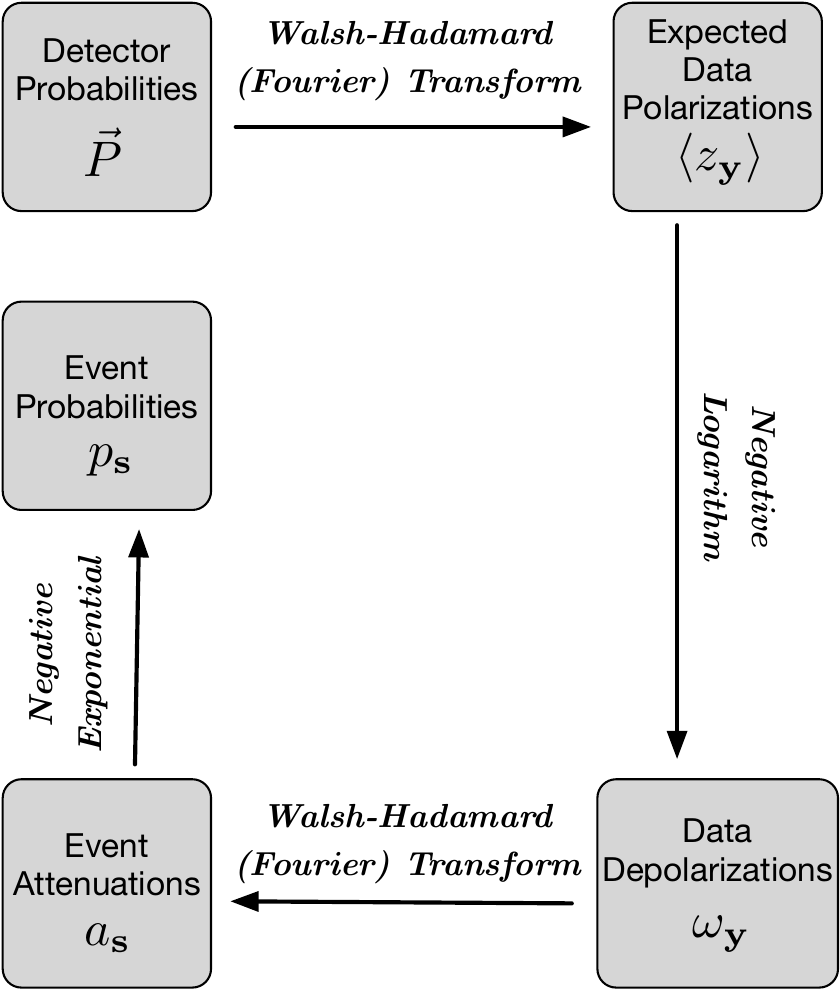}
\caption{A nontrivial chain of four transformations maps data (detector probabilities $\mathrm{Pr}(\expect{\x})$) to DEM event probabilities ($p_{\mathrm{s}}$).  There are two pairs of transformations that, if performed consecutively, would cancel out (the two Walsh-Hadamard transforms, and the negative logarithm/exponential).  They do not cancel because they are interleaved.}
\label{fig:chain}
\end{figure}

Equation \ref{eq:result1} describes an elegant relationship between observable data and DEM event probabilities, which can be used to infer DEMs from data.  But the relationship is not quite as simple as Eq.~\ref{eq:result1} suggests.  It is  a chain of steps, shown in Figure \ref{fig:chain}, some of which are packed into the notation of Eq.~\ref{eq:result1}.  We found it easy to (incorrectly) intuit that the two Walsh-Hadamard transforms -- one mapping detector probabilities $\vec{P}$ to polarizations $\expect{z_\y}$, and the other transforming depolarizations $\omega_\y$ to attenuations $a_{\s}$ -- should invert each other.  They don't because, as shown in Fig.~\ref{fig:chain}, there's a logarithm in between the two Walsh-Hadamards (which is balanced by the exponentiation that maps attenuations to probabilities).  We find it intriguing, and still slightly mysterious, that interleaving these pairs of mutually-inverse transformations performs a useful and nontrivial mapping.

\section{Estimating aggregated DEM properties} \label{sec:aggregated}

In principle, the probabilities associated with all $2^N$ possible DEM events can be estimated by (1) estimating all $2^N$ polarizations by sampling $K$ $N$-bit strings of detector data, and (2) applying Eqs.~\ref{eq:result2}-\ref{eq:result4}. In practice, two problems arise if $N$ is large. First, working with exponentially large vectors is hard. Second, for realistic DEMs with large $N$, the polarizations $\expect{z_\y}$ for arbitrary $N$-bit strings $\y$ may become statistically indistinguishable from zero, causing $\omega_{\y} = -\ln\expect{z_\y}$ to fluctuate wildly and render estimated attenuations meaningless (see Sec.~\ref{sec:largeN}).

These issues can be evaded either by focusing on coarse-grained or ``aggregated'' DEM properties instead individual DEM events' rates, or by assuming an ansatz such as sparsity. We examine estimation of aggregated properties in this section, and sparse DEM estimation in Section \ref{sec:Sparse}.

\subsection{Reproducing the $p_{ij}$ method}

We begin by reproducing some known results \cite{spitz2018adaptive,Google2021} that are often called ``the $p_{ij}$ method''.  By doing so, we answer one of the questions that motivated us to study this topic, which is ``What exactly \textit{are} the $p_{ij}$ coefficients?''  We conclude that they are \textit{aggregated} probabilities of large classes of DEM events -- specifically, of all DEM events that flip both $x_i$ and $x_j$ -- and in our notation they are written as $p_{\{i,j\}*}$.

\subsubsection{Calculating $p_{ij}$ for $N=2$}

Suppose there are only two detectors, so $N=2$.  Four detector histories can be observed: $\x\in\{\mathtt{[00],[01],[10],[11]}\}$.  There are three possible DEM events, and a completely general DEM is:

\begin{center}
\begin{tabular}{|c|c|c|}
 \hline
 \multicolumn{3}{|c|}{$N=2$ DEM} \\
 \hline
 Event ($E$) & $\s$ & Probability  \\
 \hline
 $\{1\}$ & [\texttt{10}] & $p_{\{1\}}$ \\
 $\{2\}$ & [\texttt{01}] & $p_{\{2\}}$ \\
 $\{1,2\}$ & [\texttt{11}] & $p_{\{1,2\}}$ \\
 \hline
\end{tabular}
\end{center}

Now, Eq.~\ref{eq:result3} tells us how to estimate $d_\s = 1-2p_\s$ from empirical polarizations $\expect{z_\y}$, where both events $\s$ and parities $\y$ range over $\{\mathtt{[01],[10],[11]}\}$.  In particular,
\begin{equation}
    d_{\{1,2\}} = \left(\frac{\expect{z_{\mathtt{[10]}}}\expect{z_{\mathtt{[01]}}}}{\expect{z_{\mathtt{[00]}}}\expect{z_{\mathtt{[11]}}}}\right)^{1/2} = \left(\frac{\expect{z_{\mathtt{[10]}}}\expect{z_{\mathtt{[01]}}}}{\expect{z_{\mathtt{[11]}}}}\right)^{1/2},
\end{equation}
because $z_{\mathtt{[00]}}$ is always 1.  These expressions are easier to recognize if we rewrite $\expect{z_{\mathtt{[10]}}}$, $\expect{z_{\mathtt{[01]}}}$, and $\expect{z_{\mathtt{[11]}}}$ in notation consistent with Ref.~\cite{spitz2018adaptive,Google2021}, as expectation values of single-bit parities $z_1$, $z_2$ and $z_1z_2$ respectively, to get
\begin{align}
    1-2p_{\{1,2\}} &= \left(\frac{\expect{z_2}\expect{z_1}}{\expect{z_1z_2}}\right)^{1/2} \\
    p_{\{1,2\}} &= \frac12 - \frac12\left(\frac{(1-2\expect{x_1})(1-2\expect{x_2})}{1-2\expect{x_i}-2\expect{x_i}+4\expect{x_ix_j}}\right)^{1/2}. \label{eq:GooglePij}
\end{align}
This is Equation (11) from the Supplementary Material to Ref.~\cite{Google2021}, mod a little algebra.  So we have rederived the same formula for $p_{ij}$ (or at least $p_{12}$).

\subsubsection{Solving $N>2$ with reduced DEMs}

However, we haven't derived this formula in the same context as Refs.~\cite{spitz2018adaptive, Google2021}, because they considered more than 2 detectors. We can easily extend the calculation to $N$ detectors using reduced DEMs.  For example, a DEM that models \textit{all} weight-2 events on 4 detectors is:

\begin{center}
\begin{tabular}{|c|c|}
 \hline
 \multicolumn{2}{|c|}{$N=4$ DEM} \\
 \hline
 Event ($\s$) & Probability \\
 \hline
 \texttt{[1100]} & $p_{\{1,2\}}$ \\
 \texttt{[1010]} & $p_{\{1,3\}}$ \\
 \texttt{[0110]} & $p_{\{2,3\}}$ \\
 \texttt{[1001]} & $p_{\{1,4\}}$ \\
 \texttt{[0101]} & $p_{\{2,4\}}$ \\
 \texttt{[0011]} & $p_{\{3,4\}}$ \\
 \hline
\end{tabular}
\end{center}

Now, suppose (without loss of generality) that we want to estimate $p_{\{1,2\}}$.  \texttt{[1100]} is the \textit{only} event that flips both $x_1$ and $x_2$.  Therefore, the values of $x_3$ and $x_4$ cannot be relevant for the purpose of estimating $p_{\{1,2\}}$ -- they do not detect the event \texttt{[1100]}, and they are not necessary to distinguish it from other events.  We can therefore construct a \textit{reduced DEM} supported only on $\{x_1,x_2\}$:

\begin{center}
\begin{tabular}{|c|c|}
 \hline
 \multicolumn{2}{|c|}{Reduced DEM for $\{x_1,x_2\}$} \\
 \hline
 Event ($\s$) & Probability \\
 \hline
 \texttt{[11]} & $p^{\rm r}_{\{1,2\}} \equiv p_{\{1,2\}}$  \\
 \texttt{[10]} & $p^{\rm r}_{\{1\}} \equiv p_{\{1,3\}}+p_{\{1,4\}} - 2p_{\{1,3\}}p_{\{1,4\}}$ \\
 \texttt{[01]} & $p^{\rm r}_{\{2\}} \equiv p_{\{2,3\}}+p_{\{2,4\}} - 2p_{\{2,3\}}p_{\{2,4\}}$ \\
 \hline
\end{tabular}
\end{center}

The math and solution are identical to the original $N=2$ case, and we isolate the same $p_{\{1,2\}}$ (Eq.~\ref{eq:GooglePij}).  But this analysis highlights something important: restricting attention to $k<N$ bits aggregates events from the original DEM.  For example, $p^{\rm r}_{\{1\}}$ is not just the probability of a single event that flips $x_1$.  It is the \textit{aggregated} probability of all events that flip $x_1$ but not $x_2$.  

Similarly, $p^{\rm r}_{\{1,2\}}$ is the aggregated probability of all events that flip both $x_1$ and $x_2$.  In this specific example, \texttt{[1100]} was (by construction) the only such event, so the reduced DEM isolated $p_{\{1,2\}}$. But if the original DEM had contained another event \texttt{[1111]} with probability $p_{\{1,2,3,4\}}$, then the reduced DEM would be (written, for convenience, using attenuation instead of probability):

\begin{center}
\begin{tabular}{|c|c|}
 \hline
 \multicolumn{2}{|c|}{Reduced DEM for $\{x_1,x_2\}$} \\
 \hline
 Event ($\s$) & Attenuation \\
 \hline
 \texttt{[11]} & $a^{\rm r}_{\{1,2\}} = a_{\{1,2\}} + a_{\{1,2,3,4\}}$ \\
 \texttt{[10]} & $a^{\rm r}_{\{1\}} = a_{\{1,3\}} + a_{\{1,4\}}$ \\
 \texttt{[01]} & $a^{\rm r}_{\{2\}} = a_{\{2,3\}} + a_{\{2,4\}}$ \\
 \hline
\end{tabular}
\end{center}

So in general, estimating the probability or attenuation associated with two bits -- e.g.~$a^{\rm r}_{\{1,2\}}$ -- does not reliably isolate a single event.  Instead, it aggregates the attenuation of all events that flip both bits.

It's tempting to infer that ``$a^{\rm r}_{\{1\}}$'' aggregates the attenuation of all events that flip $x_1$. It doesn't.  \textit{In this context} -- i.e., when we restrict attention to $\{x_1,x_2\}$ -- $a^{\rm r}_{\{1\}}$ is the aggregated attenuation of all events that flip $x_1$ but \textit{not} $x_2$.  But in other contexts, it would have different values:
\begin{itemize}
    \item In the context of the original $N=4$ DEM, ``$a_{\{1\}}$'' would be the attenuation of the unique event \texttt{[1000]} that flips $x_1$ but \textit{not} $x_2$, $x_3$, or $x_4$.
    \item In the context of an even-more-reduced 1-bit DEM for bit $x_1$ only, ``$a^{\rm r}_{\{1\}}$'' would be the aggregated attenuation of \textit{all} events that flip $x_1$.
\end{itemize}
We need better notation for estimatable DEM properties (e.g. probabilities, decay factors, and attenuations).  

\subsection{Describing aggregated properties}

The preceding analysis showed that easily estimatable properties are indexed by (1) a set of detectors that get flipped, (2) a set of detectors that \textit{don't} get flipped, and (3) the remaining detectors that are ignored.  We call the set of events defined by these conditions a \textit{class}.  It can be useful to estimate the aggregated probability, decay factor, or attenuation of a class.  There are at least two candidate notations for identifying classes.

First, when $N$ is well-defined, we can define classes by \textit{ternary} strings, where the 3rd symbol ``$\mathtt{x}$'' indicates a bit whose value is aggregated over, e.g.
\begin{equation}
    a_{\mathtt{[10xx]}} = a_{\mathtt{[1000]}} + a_{\mathtt{[1001]}} + a_{\mathtt{[1010]}} + a_{\mathtt{[1011]}}.
\end{equation}

Second, an equivalent convention that does not require specifying $N$ is to list both the specified indices and their values, e.g.
\begin{align}
    a_{\{1,2\}=\mathtt{[10]}} &= a_{\mathtt{[10]}}\ \mathrm{for\ }N=2 \\
        &= a_{\mathtt{[100]}} + a_{\mathtt{[101]}}\ \mathrm{for\ }N=3 \\
        &= a_{\mathtt{[1000]}} + a_{\mathtt{[1001]}} + a_{\mathtt{[1010]}} + a_{\mathtt{[1011]}}\\
        &\ldots
\end{align}
Hereafter, we will use these two conventions to define classes and index aggregated $a$, $d$, and $p$ properties, as appropriate. Because we're lazy, we will use an abbreviated notation for aggregated quantities defined by a set of bits that are all equal to 1:
\begin{equation}
    a_{E*} \equiv a_{E = \mathtt{[1^{|E|}]}}.
\end{equation}
So, for example, $a_{\{1,2\}*} = a_{\{1,2\}=\mathtt{[11]}}$ is the aggregated attenuation of the class of events that flip bits 1 and 2, and $p_{\{7\}*} = p_{\{7\}=\mathtt{[1]}}$ is the aggregated probability of the class that flips bit 7.

\subsection{Easily estimatable aggregated properties}

Aggregated properties of classes are very useful because their estimates can be computed very efficiently.  To see how this works, suppose we have $N$-bit detector data for $N\gg1$, and we want to estimate the aggregated attenuation of all DEM events that flip $x_1$, regardless of what other bits they do or don't flip.

We could construct a reduced 1-bit DEM,
\begin{center}
\begin{tabular}{|c|c|}
 \hline
 \multicolumn{2}{|c|}{Reduced DEM} \\
 \hline
 Event ($\s$) & Attenuation \\
 \hline
 \texttt{[1]} & $a^{\rm r}_{\mathtt{[1]}}$ \\
 \hline
\end{tabular}
\end{center}
then estimate $\omega_{\mathtt{[1]}} = -\ln(\expect{z_1})$, and finally use Eq.~\ref{eq:matrix} to estimate $a^{\rm r}_{\mathtt{[1]}} = \omega_{\mathtt{[1]}}$.

Equivalently, we can simply observe that in $N$-bit data, the depolarization $\omega_{\{1\}}$ of a single bit $x_1$ is equal to the aggregated attenuation of \textit{all} $N$-bit DEM events that flip bit 1 ($x_1$),
\begin{equation}
    \omega_{\{1\}} = \sum_{\s:\ s_1=1}{a_\s}.
\end{equation}
This yields the same result, but stated more transparently as $a_{\{1\}=\mathtt{[1]}} = \omega_{\{1\}}$.

This approach generalizes.  The estimated attenuation of any $N$-bit \textit{single} DEM event (Eq.~\ref{eq:matrix}) contains $2^N-1$ terms.  For example, if we have $N=3$-bit data and want to estimate the attenuation of the event \texttt{[111]}, it is
\begin{align}
    a_{\mathtt{[111]}} &= \frac14 \left(\begin{array}{l}\omega_{\mathtt{[100]}} + \omega_{\mathtt{[010]}} + \omega_{\mathtt{[001]}} + \omega_{\mathtt{[111]}} \\ 
    - \omega_{\mathtt{[110]}} - \omega_{\mathtt{[101]}} - \omega_{\mathtt{[011]}}\end{array}\right) \\
    &= \frac14\left(\begin{array}{l}\omega_{\{1\}} + \omega_{\{2\}} + \omega_{\{3\}} + \omega_{\{1,2,3\}} \\ 
    - \omega_{\{1,2\}} - \omega_{\{1,3\}} - \omega_{\{2,3\}}\end{array}\right).
\end{align}
For $N>3$, these formulae become hard to write down, and eventually become infeasible to evaluate exactly.  But if instead we compute these formulae for small $k\ll N$, they yield aggregated properties of classes.  For example, independent of $N$,
\begin{equation}
    a_{\{1,2,3\}=\mathtt{[111]}} = \frac14\left(\begin{array}{l}\omega_{\{1\}} + \omega_{\{2\}} + \omega_{\{3\}} + \omega_{\{1,2,3\}}\\ - \omega_{\{1,2\}} - \omega_{\{1,3\}} - \omega_{\{2,3\}}\end{array}\right),
\end{equation}
and
\begin{equation}
    a_{\{5,7\}=\mathtt{[10]}} = \frac12\left(\omega_{\{5\}} + \omega_{\{5,7\}} - \omega_{\{7\}}\right).
\end{equation}
These formulae allow estimating the aggregated attenuation of all events consistent with a substring (e.g. $\{5,7\}=[10]$) \textit{as long as} the substring contains at least one 1.  We do not know of an efficient exact formula for (e.g.) the aggregated attenuation of the class of events consistent with $\{5,7\}=\mathtt{[00]}$.

\subsection{Monte Carlo estimation} \label{sec:MonteCarlo}

Estimators of DEM properties are functions of the observable depolarizations $\vec{\omega}$.  When the DEM property involves relatively few detector bits -- e.g., if $N$ is small, or we're estimating aggregated attenuations associated with a $k$-bit subset of detectors and $k$ is small -- we can compute these functions exactly.

But some useful estimators, like the attenuation of a unique DEM event on $N\gg1$ bits (Eq.~\ref{eq:result2}), are linear combinations of exponentially many distinct depolarizations.  Computing these functions exactly would require too much time.  In these cases, we can compute approximate estimators by Monte Carlo sampling instead.  For example, $a_\s$ can be estimated by writing Eq.~\ref{eq:result2} as
\begin{equation*}
    a_\s = 2 \expect{\left[ {(-1)^{\y\cdot\s+1}\omega_{\y}}\right]}_\y,
\end{equation*}
and then approximating the average over all $2^N$ strings $\y$ (\textit{including} [$0^N$]) by drawing a small random sample of $N$-bit strings and averaging over them.

A relevant example of a quantity that can be estimated this way (and for which we know no other estimator) is the total attenuation of \textit{all} DEM events.  This quantity, which we denote $a_0$, is a useful measure of the \textit{total} noise observed in syndrome extraction.  It is given by
\begin{align}
    a_0 &= \sum_\s{a_\s} \nonumber \\
        &= -2^{1-N/2}(\vec{1}-\vec{\mathbf{0}})^T H \vec{\omega} \nonumber \\
        &= 2 (2^{-N}\vec{1} - \vec{\mathbf{0}})^T \vec{\omega} \nonumber \\
        &= 2\expect{ \omega_\y }_\y, \label{eq:total}
\end{align}
where again the average is over all $2^N$ strings $\omega_{\y}$ including $[0^N]$. This formula has a simple intuitive explanation; if $\y$ is chosen uniformly at random, then the polarization $\expect{z_{\y}}$ will be affected by (on average) half the DEM events, so we can capture all of them by averaging over $\y$ and doubling.  Evaluating it exactly requires computing $2^N-1$ polarizations, but it can be efficiently approximated using Monte Carlo.

\subsubsection{Divergent depolarizations for large $N$} \label{sec:largeN}

These Monte Carlo algorithms have an important Achilles heel: they require random sampling over \textit{all} strings $\y$. As mentioned at the beginning of Sec.~\ref{sec:aggregated}, polarizations $\expect{z_{\y}}$ associated with high-weight $\y$ can be statistically indistinguishable from zero, causing very large fluctuations in the associated $\omega_{\y}$.  We now unpack that assertion.

So far, we have avoided making assumptions about the form of DEMs -- i.e., about which events $\mathbf{s}$ have probabilities that are large or nonzero -- because we wanted to make the most broadly applicable statements.  However, the DEMs that describe real-world syndrome extraction data are \textit{not} expected to be even remotely generic.  

A useful example of what we do \textit{not} expect to see in real-world QEC is the ``uniformly depolarizing DEM'', in which \textit{every} $N$-bit event $\mathbf{s}$ occurs with probability $\epsilon/2^N$.  With probability $(1-\epsilon/2^N)^{2^N} \approx e^{-\epsilon} = 1-\epsilon+O(\epsilon^2)$, no event happens.  When an event does occur, with probability $1-e^{-\epsilon} = \epsilon + O(\epsilon^2)$, it is a uniformly random $\mathbf{s}$. Therefore, every parity's depolarization $\omega_{\mathbf{y}}$ is identically given by (Eq.~\ref{eq:omega}) $\omega_{\mathbf{y}} = -2^{N-1}\ln(1-\epsilon/2^{N-1}) \approx \epsilon$.

In contrast, realistic DEMs will be dominated by events that are low-weight and local -- i.e., strings $\mathbf{s}$ in which only a few bits are 1, and those bits are relatively close to each other in time and space.  Exceptions will exist -- e.g., frequency crosstalk can cause distant bits to flip together, and catastrophic events like cosmic rays may flip many detectors -- but they should be rare and contribute a small fraction of the DEM's total attenuation.  Furthermore, we expect realistic DEMs to be approximately \textit{extensive}, meaning that the aggregated probability of any single detector flipping is $O(\epsilon)$ for some $\epsilon \ll 1$, and the total attenuation of an $N$-bit DEM grows linearly with $N$ as $O(\epsilon N)$.

It is therefore reasonable to expect that $\omega_\y$ will grow approximately linearly with the Hamming weight of $\y$.  (This is \textit{precisely} true if the DEM contains all weight-1 events $\s$ with equal probabilities).  Low-weight depolarizations $\omega_\y$ can be estimated reliably from data, but when $\y$'s Hamming weight exceeds some threshold $w_{\mathrm{max}}$, the fluctuations in $\omega_\y$ become large enough to prevent its use for estimating the DEM.

There are at least three ways to avoid this problem.  One is to focus exclusively on aggregated properties, as we did above.  These can be estimated from low-weight polarizations.  A second is to simply restrict attention to subsets of $k \leq w_{\mathrm{max}}$ detector bits, using reduced DEMs, and accept that correlations between those bits and the remainder will not be correctly identified.  The third, which we find most interesting, is to assume that the DEM has some structure -- i.e., it is sparse -- and leverage that structure to estimate individual DEM events' attenuations using only low-weight polarizations.  In the next section, we explore some specific cases of this approach.

\section{Estimating sparse DEMs} \label{sec:Sparse}

An $N$-bit DEM is \textit{sparse} if the probabilities of all but $\mathrm{poly}(N)$ of the $2^N$ possible events $\s$ can be approximated, with negligible consequences, as zero.  In practice, we will consider the simpler condition of ``all but $\mathrm{poly}(N)$ event probabilities \textit{are} zero,'' but since this condition isn't physically verifiable, it's useful to keep the true definition in mind.

Sparsity is a powerful, and almost necessary, assumption.  If $2^N$ is intractably large, then estimating a non-sparse DEM to high precision is unlikely to be feasible (if only because just writing down all the event probabilities would take a long time).  But if the DEM is sparse, then Eq.~\ref{eq:result1} can be solved (in principle) using sparse Walsh-Hadamard transform algorithms \cite{hamood2011fast, li2014spright, chen2015robust, cheraghchi2017nearly}, which are special cases of compressed sensing \cite{donoho2006compressed} and sparse Fourier transforms \cite{rajaby2022structured}.  These algorithms are remarkable in that, given \textit{only} the sparsity assumption and no prior information on \textit{which} $\mathrm{poly}(N)$ events have nonzero probabilities, they can identify the events with nonzero probabilities as well as their probabilities.

The algorithms we consider here are more modest.  We start with the (not unreasonable) assumption that we know a $\mathrm{poly}(N)$-sized set of candidate events $\s$, and we want to estimate their attenuations.  We propose two algorithms, the second of which \textit{may} in practice provide some or all of the functionality of the more general sparse Walsh-Hadamard algorithms.  We defer the work of demonstrating and analyzing these algorithms to future work; our goal here is simply to show that the divergent depolarizations problem discussed in the previous section is not an insurmountable obstacle to practical estimation of DEMs from data.

\subsection{Inverting submatrices of $H$}

Let's assume that the only depolarizations $\omega_\y$ that can be reliably estimated from data are those indexed by strings $\y$ with Hamming weight $\leq w_\mathrm{max}$.  There are $D \equiv \binom{N+w_\mathrm{max}-1}{w_\mathrm{max}} \approx N^{w_\mathrm{max}}/w_\mathrm{max}!$ of these\footnote{Summing the $\binom{N}{w}$ strings of Hamming weight $w$ over $w=1\ldots w_{\mathrm{max}}$ yields $\binom{N+w_\mathrm{max}-1}{w_\mathrm{max}}$.} (assuming $w_\mathrm{max} \ll N$).  Even if $2^N$-dimensional linear algebra is feasible, we do not have enough data to uniquely infer all $2^N$ DEM event attenuations.

However, we do have enough data to (in principle) infer the attenuations of a set $\{\s_1,\ldots \s_L\}$ of $L\leq D$ distinct events.  Doing so is easy.  Beginning with Eq.\ref{eq:matrix},
\begin{equation*}
    \vec{\omega} = W\vec{a},\ \mathrm{where}\ W_{\y,\s} = \y\cdot\s,
\end{equation*}
we carve out a submatrix $W'$ of $W$ indexed by (1) the $D$ rows corresponding to low-weight strings $\y$, and (2) the $L$ columns corresponding to $\{\s_1,\ldots \s_L\}$.  Assuming attenuations of all other events are zero, it follows that
\begin{equation}
    \vec{\omega'} = W'\vec{a'}, \label{eq:submatrix}
\end{equation}
where $\vec{\omega'}$ and $\vec{a'}$ are the appropriate subvectors.  Now we can either invert or pseudoinvert $W'$ (depending on whether $L < D$) to get an estimate
\begin{equation}
    \vec{a'} = W'^{-1}\vec{\omega'}.
\end{equation}
If $L$ is not too big, then brute-force inversion is feasible for any chosen set $\{\s_1,\ldots \s_L\}$.  But $L$ can be as large as $\binom{N+w_\mathrm{max}-1}{w_\mathrm{max}}$, which can easily be large enough to make storing and inverting $W'$ infeasible.  

For at least one natural set of events, there is a closed-form algorithm to invert Eq.~\ref{eq:submatrix}.  This is the set of all $\s$ with Hamming weight $\leq w_{\mathrm{max}}$.  It is a natural choice; we are inferring the attenuations of $D$ low-weight events from the depolarizations of $D$ low-weight parities.

The algorithm is as follows.  Let $E$ be a set of $|E| \leq w_{\mathrm{max}}$ detector bits, e.g. $|E|=3$ for $E=\{1,4,7\}$.  To infer $a_E$:
\begin{enumerate}
\item If $|E| = w_{\mathrm{max}}$, then $E$ is the \emph{only} event with weight $\leq w_{\mathrm{max}}$ that flips all the bits in $E$.  Therefore, its attenuation is equal to the aggregated attenuation $a_{E*}$, which can be computed using Eq.~\ref{eq:result2} as described in Sec.~\ref{sec:aggregated}.
\item If $|E| < w_{\mathrm{max}}$, then $E$ is the only event with weight $\leq |E|$ that flips all the bits in $E$, but there are $\binom{N+w_{\mathrm{max}}-2|E|-1}{w_{\mathrm{max}}-|E|}$ additional events\footnote{Given a set of $|E|$ bits, there are $N-|E|$ other bits.  Adding any subset containing $w=1\ldots w_{\mathrm{max}}-|E|$ of them yields one of the events being counted.  There are $\binom{N-|E|}{w}$ such subsets of size $w$.  Summing this over $w=1\ldots w_{\mathrm{max}}-|E|$ yields the given expression.} with weights in $(|E|,w_{\mathrm{max}}]$ that also flip all the bits in $E$.  Subtracting all of those events' attenuation from the aggregated attenuation $a_{E*}$ yields $a_{E}$.
\end{enumerate}
These rules can be used to compute all $D$ low-weight attenuations in a relatively time-efficient manner by computing the highest-weight attenuations first, and using their already-computed values to perform the subtraction efficiently.  If the $O(D)$ space required for this algorithm is prohibitive, they can also be used to compute any individual attenuation via a space-efficient recursive algorithm that computes aggregated attenuations on the fly without caching them.  It seems likely that clever caching could yield space-efficient algorithms with better time complexity.

\subsection{Estimating low-weight DEMs by pruning the class lattice} \label{sec:lattice}

\begin{figure}[t!]
\centering
\includegraphics[width=8cm]{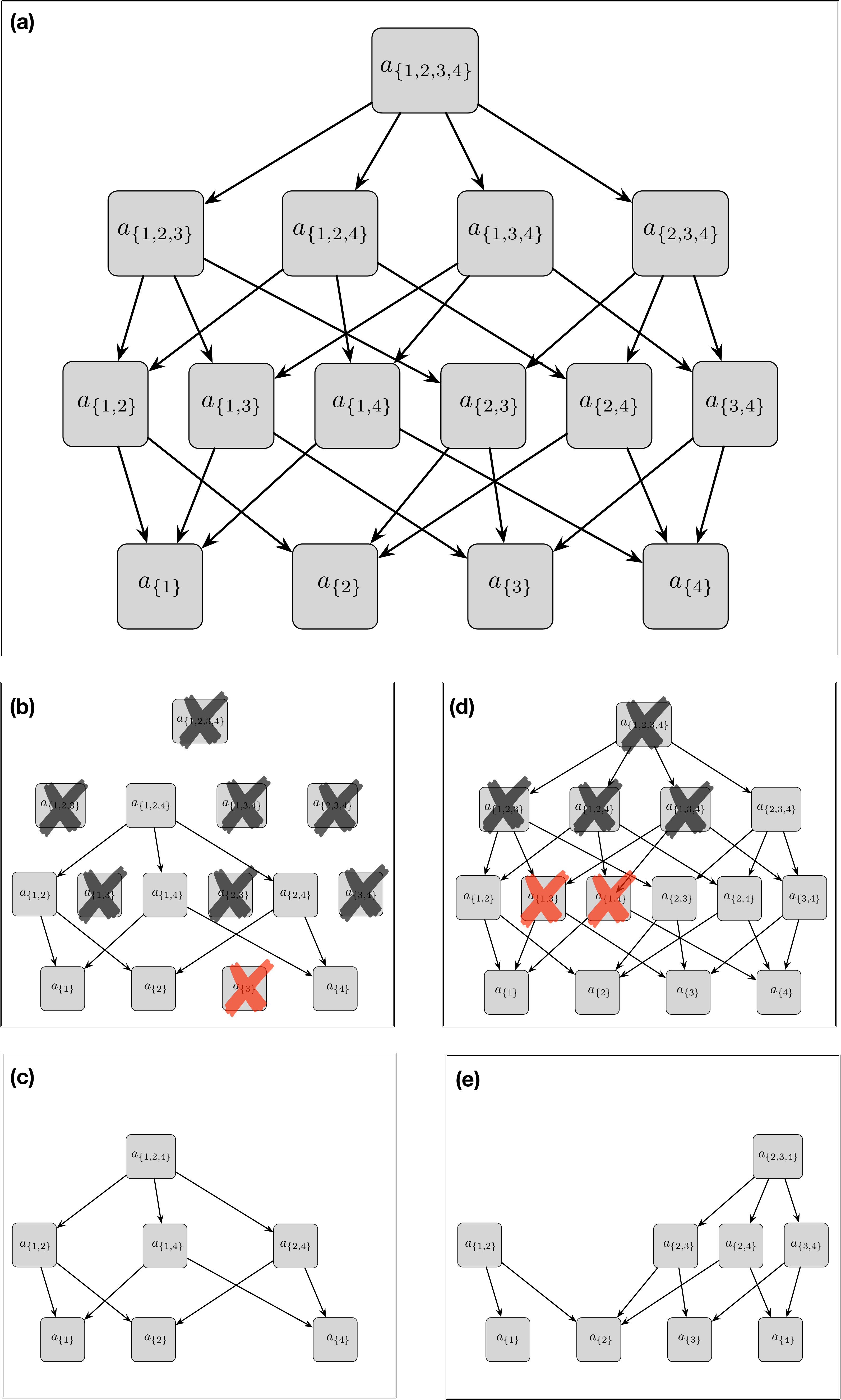}
\caption{Classes of events whose aggregated attenuation can be easily estimated form a lattice, shown for $N=4$ in panel (a).  Arrows indicate inclusion, e.g. the class of events that flip $x_1$ and $x_2$ is a subset of the class that flips $x_1$.  The algorithm presented in Sec.~\ref{sec:lattice} starts by estimating attenuations of large classes at the bottom of the lattice.  It moves upward, subdividing classes.  When a class is found to be empty (red X) -- i.e., its attenuation is indistinguishable from zero, as illustrated for class $\{3\}$ in panel (b) and classes $\{1,3\}$ and $\{1,4\}$ in panel (d) -- the lattice can be pruned by removing all the subclasses (black X) of the empty class, as shown in panels (c) and (e).}
\label{fig:lattice}
\end{figure}

The preceding algorithm does not fully leverage sparsity, because it computes \emph{all} the attenuations for events with weights $1,2,\ldots w_{\mathrm{max}}$.  This set grows very rapidly with $w_{\mathrm{max}}$ -- e.g., for $N=100$ and $w_{\mathrm{max}}=8$, there are about $3.2\times 10^{11}$ events.  In any real-world experiment, the overwhelming majority of these will not occur.  Identifying the few that do is like searching for a needle in a haystack.

We can do so more efficiently with an algorithm that computes aggregated attenuations, starting with those of large classes, then progressively subdivides those classes until they contain just one event.  This algorithm explores the \textit{lattice} of event classes, shown in Fig.~\ref{fig:lattice}a, ordered by inclusion.  For example, the $2^2$-event classes $\{1,2\}$, $\{1,2\}$, $\{1,2\}$, and $\{1,2\}$ are all subsets of the $2^3$-event class $\{1\}$.  The bottom of the lattice contains relatively few classes -- just $N$ large classes in the first layer, $\binom{N}{2}$ in the next layer, etc -- and their aggregated attenuations can be computed very efficiently.

Computing \textit{every} attenuation in the bottom $w_{\mathrm{max}}$ layers of the class lattice is just as arduous as the algorithm described in the previous section.  However, if the unknown DEM is both sparse and low-weight, this isn't necessary.  The key is that, as the algorithm works its way up the lattice (estimating and storing aggregated attenuations), whenever it finds a class whose aggregated attenuation is statistically indistinguishable from zero, it can ``prune'' every subclass of the empty class (since they must be empty as well).  This is illustrated in Fig.~\ref{fig:lattice}b-c (where the class $\{3\}$ is empty) and Fig.~\ref{fig:lattice}d-e (where classes $\{1,3\}$ and $\{1,4\}$ are empty).

Pruning enables remarkable savings for sparse low-weight DEMs.  An event of weight $w$ contributes to $2^w-1$ classes.  If a DEM contains $L$ events, and their weights are at most $w_{\mathrm{max}}$, then at \textit{most} $L(2^w-1)$ classes can have nonzero attenuation.  If we consider an $N=100$-bit DEM with $L=1000$ events of weight at most $w_{\mathrm{max}}=8$, at most $2.55\times 10^5$ classes (out of $3.2\times 10^{11}$) have nonzero attenuation, and in realistic cases there will be far fewer.

The algorithm is as follows:
\begin{enumerate}
\item Compute $a_{\{i\}*}$ for each class that flips a single detector bit.  If $a_{\{i\}*}=0$ for any bit $x_i$, then no event's support includes $x_i$, and it can be excluded from all further analysis.  (This is very unlikely to happen, but it illustrates a principle used repeatedly in this algorithm).  Store a list of all bits $i$ for which $a_{\{i\}*}\neq 0$, and their associated attenuations.
\item For each pair of detector bits $\{i,j\}$ that were not excluded in the previous step, compute $a_{\{i,j\}*}$.  Store a list of all pairs for which $a_{\{i,j\}*}\neq 0$, and their associated attenuations.  We call this the \textit{weight-2 decoder graph}.
\item Now, if a weight-3 class $\{i,j,k\}$ has nonzero attenuation $a_{\{i,j,k\}*} > 0$, then all three weight-2 classes $\{i,j\}$, $\{j,k\}$, and $\{i,k\}$ must have attenuations at least as large. Therefore, the only weight-3 classes that can have nonzero attenuation are those corresponding to 3-element \textit{cliques} in the weight-2 decoder graph.  All others can be excluded.
\item For each 3-element clique $\{i,j,k\}$ compute $a_{\{i,j,k\}*}$. Store a list of all triples $\{i,j,k\}$ with nonzero attenuation.  We call this the \textit{weight-3 decoder hypergraph}.  Now, applying the same reasoning again, a weight-4 class can only have nonzero attenuation if it corresponds to a 4-element hyperclique (a set of 4 elements that are fully connected by 3-element hyperedges) in the weight-3 decoder hypergraph.
\item For each 4-element hyperclique $\{i,j,k,l\}$, compute $a_{\{i,j,k,l\}*}$, store a list of all the ones with nonzero attenuation, etc.  Continue to iterate up the lattice, identifying successfully larger hypercliques and excluding all $w$-tuples that do not form hypercliques, until no nonempty classes are found (e.g., at level $w_{\mathrm{max}}$, if not before).
\end{enumerate}
The first two steps are identical to the $p_{ij}$ algorithm used (e.g.) in Ref.~\cite{Google2021}, which can be used to construct a data-driven decoder graph for matching decoders.  But instead of stopping there, and interpreting each nonzero $p_{ij}$ (or $a_{\{i,j\}*}$) as a weight-2 DEM event (corresponding to an edge in the decoder graph), this algorithm checks to see whether any of a 2-bit class's attenuation is actually attributable to higher-weight DEM events, by (1) looking for 3-element cliques, and (2) testing to see whether there is evidence for a weight-3 or higher DEM event.  Note that cliques do not always indicate a higher-weight event -- it's entirely possible that $\{i,j\}$, $\{j,k\}$, and $\{i,k\}$ are all independent weight-2 DEM events, and if they are then the algorithm will find $a_{\{i,j,k\}*}=0$.

Once the entire pruned lattice has been constructed (as shown in Fig.~\ref{fig:lattice}c,e), it is straightforward to compute attenuations of individual DEM events as follows:
\begin{enumerate}
    \item Every class $C$ that has no subclasses above it (e.g.~$\{2,3,4\}$ and $\{1,2\}$ in Fig.~\ref{fig:lattice}e) represents a unique event.  Record it and its attenuation in the estimated DEM.  Subtract its attenuation from all of its superclasses (i.e., those below it in the pruned lattice). Any class whose attenuation is reduced to zero by this subtraction should be pruned from the lattice.  Finally, remove $C$ from the lattice.
    \item Repeat step 1 with the newly pruned lattice, until no more classes are left.
\end{enumerate}

We expect that more, and better, algorithms can be constructed for inferring DEMs and interesting aggregated properties of them from data.  Our primary goal in this paper was to lay out a framework for doing so.  We believe that the specific algorithms outlined above serve at least to demonstrate feasibility of the approach (and motivate further investigation), and may be useful in their own right, especially if augmented with statistical significance testing to identify when an estimated attenuation is statistically significant (or, conversely, statistically indistinguishable from zero).

\section*{Acknowledgements}

This article was authored by employees of National Technology \& Engineering Solutions of Sandia, LLC under Contract No.~DE-NA0003525 with the U.S.~Department of Energy (DOE). The employees own all right, title, and interest in and to the article and are solely responsible for its contents. The U.S.~Government retains, and the publisher, by accepting the article for publication, acknowledges that the U.S.~Government retains, a non-exclusive, paid-up, irrevocable, worldwide license to publish or reproduce the published form of this article or allow others to do so, for U.S.~Government purposes. DOE will provide public access to these results of federally sponsored research in accordance with the DOE Public Access Plan \href{https://www.energy.gov/downloads/doe-public-access-plan}{https://www.energy.gov/downloads/doe-public-access-plan}.  This paper describes objective technical results and analysis. Any subjective views or opinions expressed in the paper do not necessarily represent the views of the U.S.~Department of Energy or the U.S.~Government.

\bibliographystyle{quantum}
\bibliography{DEM}
\end{document}